\documentclass[a4paper,twosides]{article}
\usepackage{CJK,multicol,multirow,graphics,fancyhdr,epstopdf}
\usepackage{amsmath,amsfonts,amssymb,bm,upgreek,mathrsfs,ccmap,mathcomp}
\usepackage[pagewise,switch,columnwise]{lineno}
\usepackage[compress,nospace]{cite}
\usepackage[dvipsnames]{xcolor}
\usepackage{CPL}

\usepackage{booktabs}
\usepackage{graphicx} 
\usepackage{tabularx} 

\newcommand{\cplyear}{2024} \newcommand{\cplvol}{Vol. 41, No. 6}
 \newcommand{\cplpagenumber}{066101}
 \newcommand{\cplpage}{\cplpagenumber-\thepage}

\begin{document}

\begin{CJK}{GBK}{song}\vspace* {-4mm} \begin{center}
\large\bf{\boldmath{Interatomic Interaction Models for Magnetic Materials: Recent Advances}}
\footnotetext{\hspace*{-5.4mm}$^{*}$Corresponding author. Email: ivan.novikov0590@gmail.com

\noindent\copyright\,{\cplyear}
\href{http://www.cps-net.org.cn}{Chinese Physical Society} and
\href{http://www.iop.org}{IOP Publishing Ltd}}
\\[5mm]
\normalsize \rm{}Tatiana S. Kostiuchenko$^{1}$, Alexander V. Shapeev$^{2}$, and Ivan S. Novikov$^{1,*}$
\\[3mm]\small\sl $^{1}$Emanuel Institute of Biochemical Physics RAS, 4 Kosygin Street, Moscow, 119334, Russian Federation
\\[3mm]\small\sl $^{2}$Independent investigator

\end{center}
\end{CJK}
\vskip 1.5mm

\small{\narrower Atomistic modeling is a widely employed theoretical method of computational materials science. It has found particular utility in the study of magnetic materials. Initially, magnetic empirical interatomic potentials or spin-polarized density functional theory (DFT) served as the primary models for describing interatomic interactions in atomistic simulations of magnetic systems. Furthermore, in recent years, a new class of interatomic potentials known as magnetic machine-learning interatomic potentials (magnetic MLIPs) has emerged. These MLIPs combine the computational efficiency, in terms of CPU time, of empirical potentials with the accuracy of DFT calculations. 
In this review, our focus lies on providing a comprehensive summary of the interatomic interaction models developed specifically for investigating magnetic materials. We also delve into the various problem classes to which these models can be applied. Finally, we offer insights into the future prospects of interatomic interaction model development for the exploration of magnetic materials.

\par}\vskip 3mm
\normalsize\noindent{\narrower{PACS: 61.50.Ah Theory of crystal structure, crystal symmetry; calculations and modeling; 75.47.Lx Magnetic oxides; 75.47.Np Metals and alloys; 82.20.Wt Computational modeling; simulation}}\\
\noindent{\narrower{DOI: 10.1088/0256-307X/41/6/066101}

\par}\vskip 5mm

{\it 1. Introduction.} Computational modeling is being increasingly utilized in materials science, serving as a valuable tool for advancing research. Among the various computational methods, atomistic modeling, such as molecular dynamics, has gained significant traction. The main objective of atomistic modeling in materials science is to accurately predict the physical and mechanical properties of materials, thus providing a precise representation of their behavior. The success of atomistic modeling depends on the computational efficiency, in terms of the CPU time, as well as the accuracy of interatomic interaction models used to calculate forces and energies between atoms in a given structure. Two commonly used types of interatomic interaction models in materials science are empirical potentials and Density Functional Theory (DFT)\cite{hohenberg_kohn_1964, kohn_sham_1965}. Empirical potentials offer high computational efficiency, making it possible to study systems comprising billions of atoms. However, their accuracy may not be sufficient, and they have limited applicability, primarily suited for specific materials and specific conditions. For instance, the embedded atom model (EAM) \cite{daw1983_eam} is commonly employed for studying metals and alloys, while the Tersoff potential \cite{tersoff1988} is often utilized for semiconductors and insulators. Additionally, the ReaxFF potential \cite{van2001_reaxff} is well-suited for investigating molecular systems. As opposed to empirical potentials, DFT calculations provide a higher level of accuracy in predicting material properties and have the advantage of universality across various materials. However, DFT calculations are computationally demanding and are usually limited to systems containing only a few hundred atoms. This compromise makes DFT unable to account for physical effects observed in larger systems.

\begin{figure}[h!]
    \centering
    \includegraphics[width=\textwidth]{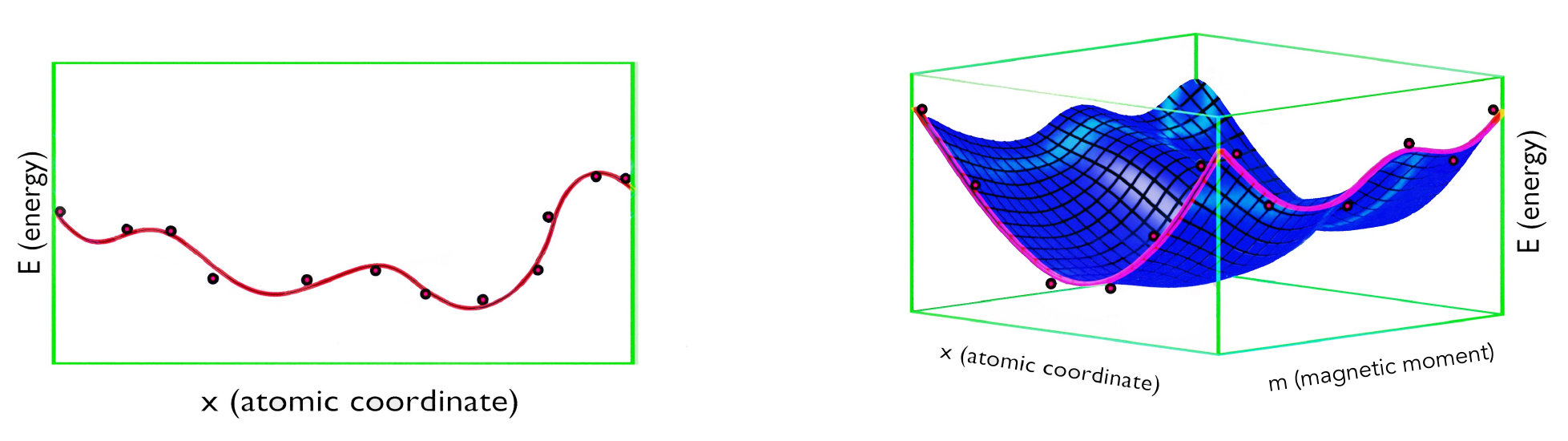}
    \caption{Illustration of the additional complexity involved in representing magnetic interatomic interaction models compared to non-magnetic models. The left side illustrates a non-magnetic model including only atomic coordinates. Because different points (configurations) may correspond to different magnetic states, the resulting potential energy function may not be a smooth function of atomic coordinates. In contrast, the right side depicts a magnetic model, incorporating both atomic coordinates and magnetic moments. The corresponding potential energy surface is described as a smooth function of the extended degrees of freedom and can thus be easily learned by a magnetic machine-learned potential.}\label{fig:degrees_of_freedom}
\end{figure}

Investigation of magnetic materials properties is of a high importance in the area of modern materials design as they are extensively utilized in various industries as structural materials and particularly in medical equipment. To accurately describe magnetic materials, including predicting the Curie temperature, it is essential to explicitly incorporate the magnetic moments of atoms in the functional form of interatomic interaction models. Representing magnetic interatomic interaction models poses greater challenges compared to non-magnetic models due to the additional degree of freedom. In a system with $N$ atoms, a non-magnetic model encompasses $3 \times N$ degrees of freedom, whereas a magnetic model necessitates consideration of $6 \times N$ degrees of freedom. We illustrate this in Fig.~\ref{fig:degrees_of_freedom}. Some empirical potentials can be applied to magnetic materials. Many of them are based on the Heisenberg model \cite{heisenberg1985theorie}, explicitly including magnetic moments in their functional form. Additionally, some magnetic empirical potentials are derived from classical non-magnetic empirical potentials, like the EAM model. However, magnetic empirical potentials share similar limitations with non-magnetic potentials: they lack universality, transferability, and may not achieve the desired level of accuracy. For precise investigations of various magnetic materials, the spin-polarized DFT calculations can be employed. These calculations allow for the exploration of different magnetic states in materials, such as ferromagnetic, antiferromagnetic, and paramagnetic properties. However, spin-polarized DFT is even more computationally demanding than non-magnetic DFT. Given the multitude of magnetic states and the computational expense of spin-polarized DFT, a substantial amount of computational resources is necessary, even for studying prototype magnetic systems like Fe. To overcome these challenges, machine-learning interatomic potentials (MLIPs) were devised as an alternative to empirical potentials and DFT. MLIPs combine the computational efficiency of empirical potentials with the accuracy and universality of DFT calculations. The pioneering work of Behler and Parrinello \cite{behler_generalized_2007} introduced MLIPs, revolutionizing the field of atomistic modeling. It is worth noting that, up until the early 2020s, conventional MLIPs did not incorporate magnetic moments in their functional form. Consequently, they were unable to address a significant class of magnetic materials. However, in recent years, various magnetic MLIPs have been developed and recommended themselves as a promising tool for investigating complex magnetic materials.

In this review, we provide a comprehensive overview of magnetic interatomic interaction models. We begin by discussing empirical potentials, followed by an exploration of spin-polarized Density Functional Theory (DFT) calculations. Additionally, we highlight the recent advancements in the development of magnetic machine-learning interatomic potentials (MLIPs). Furthermore, we address the future prospects and potential directions for the continued development of magnetic interatomic models.

{\it 2. Empirical potentials.} The interest in theoretical understanding the behavior of magnetic materials through theoretical explanations experienced a significant surge in the early 20th century. One of the pioneering atomistic-level models that emerged during this period was the Ising model, proposed by Wilhelm Lenz and his student Ernst Ising in the 1920s. This model provided a framework for investigating the magnetic states and phase transitions of materials.
In the Ising model, the material is represented by a crystalline lattice that is periodic in space, with nodes labeled by $k$. Each node is occupied by an atom that can exhibit two possible spin states: $s_i = -1$ or $s_i = 1$. The Hamiltonian function in this model can be expressed as follows:

\begin{equation} \label{Ising}
    H(s) = \sum_{\langle i j \rangle} J_{ij} s_i s_j - \mu \sum_j h_j s_j,
\end{equation}
where ${s}={s_1,\ldots,s_k}$ represents the atomic spin configuration, $\mu$ denotes the magnetic moment of a node, $J_{ij}$ is the coefficient of interaction between neighboring atoms $i$ and $j$, $h_j$ represents the external magnetic field. The first term in \eqref{Ising} is over all pairs $\langle i j \rangle$ of atoms, where $i,j=1,\ldots,N$ and $i \neq j$, $N$ is a total number of atoms in the system, the second term in \eqref{Ising} is over all atoms $j=1,\ldots,N$. The similar notation for the pairs of atoms is used in the equations below. It is common to neglect the second term, assuming the absence of an external magnetic field. Positive values of $J_{ij}$ correspond to ferromagnetic interactions, while negative values correspond to antiferromagnetic interactions. Although the one- and two-dimensional Ising models have analytical solutions \cite{ising1925contribution, peierls1936ising}, the three-dimensional case remains unresolved. Nevertheless, this model continues to be widely used as a tool for investigating phase transitions through Monte Carlo simulations \cite{elden2022monte, Benyoussef_2020}.

In less than a decade, a logical ``extension" of the Ising formalism known as the classical Heisenberg model was proposed. This model, named after the ferromagnetic formalism proposed by Heisenberg \cite{heisenberg1985theorie}, lies between a full description of the electronic state and conventional micromagnetism \cite{Nowak205}. For a one-element symmetric system, the classical Heisenberg model is expressed as:

\begin{equation}
    H = -\sum_{\langle ij \rangle} J_{ij}\bm{s}_i\cdot\bm{s}_j,
\end{equation}
where $\bm{s}_i$ and $\bm{s}_j$ are three-dimensional spins expressed as unit vectors. The specific coefficient $J_{ij}$ can be treated as an interaction constant. In the case of a system consisting of a single chemical element with all atomic positions considered equivalent, the Curie temperature of the system is approximately proportional to $J$. At temperatures much higher than $J$ (i.e., $T \gg J$), the system exists in a paramagnetic state \cite{landau2013course}. The classical Heisenberg model is commonly utilized for studying thermal effects, such as estimating the Curie temperature. For example, in the work \cite{maclaren1999electronic},  several approaches were validated for estimating the Curie temperature in ordered/disordered systems of Fe-Co alloys. 
 
 Numerous extensions of the classical Heisenberg model have been developed, and one notable variant is the Heisenberg-Landau model, originally introduced by Landau \cite{landau2013course}. This model has shown to be highly effective and is expressed as follows:
\begin{equation}
    H = -\sum_{\langle ij \rangle} J_{ij}\bm{s}_i\cdot\bm{s}_j + \sum_{i} (A\bm{s}_i^2 + B\bm{s}_i^4 + C\bm{s}_i^6).
\end{equation}
The additional term in this equation accounts for local longitudinal fluctuations, which are closely related to the local band structure and play a significant role in investigating the ferromagnetic-paramagnetic transition. Consequently, this term is commonly included in spin dynamics simulations, as demonstrated in the work by Ma et al. \cite{ma2012longitudinal}. Their study illustrated that considering longitudinal magnetic fluctuations is crucial for accurately estimating the Curie temperature. The Landau extension allows for the inclusion of higher-order terms and even angular dependence. Notably, the mean field theory based on Landau's model is more general compared to the Stoner model \cite{stoner1948mechanism}, as shown in the study of FeRh by Derlet et al. \cite{derlet2012landau}. The high accuracy of the Landau model enables its use as an alternative to computationally expensive quantum-mechanical calculations. It can generate reliable datasets, which are valuable for validating newly developed interatomic potentials, as demonstrated in the work by Domina et al. \cite{domina2022_spinSNAP}.

Another relatively simple variation of the Heisenberg model is based on the model of Stoner and Wohlfart (proposed in 1984), which initially described hysteresis effects in single-domain ferromagnets \cite{stoner1948mechanism}. A subsequent study by Rosengaard and Johannsson \cite{rosengaard1997finite} suggested introducing the Heisenberg-Hamiltonian part into the Wohlfart-Stoner model. This modification enables the description of energy changes associated with deviations from the ferromagnetic state and draws an analogy to the Ginzburg-Landau expansion:
\begin{equation}\label{Wohlfart-Stoner}
E = - \sum_{\langle i j \rangle} J_{ij} \bm{e}_i \cdot \bm{e}_j+\sum_{\langle i j \rangle}K_{ij} \left(\bm{e}_i\cdot \bm{e}_j\right)^2,
\end{equation}
where 
\begin{equation}\label{Wohlfart-Stoner_add}
J_{ij} = \sum_{k=1}^n J_{ij}^k\langle m_{ij}\rangle^{2k},\ K_{ij} = \sum_{k=1}^n K_{ij}^k\langle m_{ij}\rangle^{2k}. 
\end{equation}
The two terms in \eqref{Wohlfart-Stoner} correspond to the expansion coefficients in the Heisenberg part and biquadratic terms, respectively. Here, $\bm{e}_i$ and $\bm{e}_j$ represent unit vectors that are parallel to the local magnetic moments, and $\langle m_{ij}\rangle$ denotes the average local moment between sites $i$ and $j$. The sums in \eqref{Wohlfart-Stoner_add} are over even powers $k=2,\ldots,2n$ of magnetization expansion. The inverse correlation length, Curie temperature, lattice constant, and zero-temperature magnetization predicted with \eqref{Wohlfart-Stoner} were found to be in good agreement with experimental data for pure Fe, Co, and Ni \cite{rosengaard1997finite}.

We note, there exist semi-empirical magnetic models that are derived from the underlying semi-empirical non-magnetic models. One of these models was proposed by Dudarev and Derlet in their work \cite{dudarev2005magnetic} where they investigated defects in bcc Fe. This model closely resembles the embedded atom model (EAM) and can be expressed as follows:
\begin{equation} \label{EAM_Dudarev}
    E = \sum_i F(\rho_i)+\frac{1}{2}\sum_{\langle ij\rangle}V(r_{ij}),
\end{equation}
where the embedding function $F$ depending on the electronic density $\rho_i$ associated with the $i$-th atom has the form:
\begin{equation} \label{EAM_Dudarev_F}
    F(\rho_i)=-A\sqrt{\rho_i}-B\frac{1}{\ln{2}}(1-\sqrt{\rho_i})\ln{(2-\rho_i)}\Theta(1-\rho_i) ~{\rm{with}}~ \rho_i = \sum_{j \neq i} f(r_{ij}).
\end{equation}
In \eqref{EAM_Dudarev}, $r_{ij}$ represents the interatomic distance, while $V$ and $f$ are the radial functions that undergo parameterization. The first term in the equation \eqref{EAM_Dudarev_F} is associated with non-magnetic interactions and accounts for a symmetry-breaking band magnetism model, enabling the description of ground state properties of 3d-transition metals. The second term in \eqref{EAM_Dudarev_F} describes magnetic interactions and captures second-order phase transitions and the emergence of spontaneous magnetization. The Heaviside function, denoted as $\Theta(1-\rho_i)$, is defined as zero when $\rho_i > 1$ and one for all other cases. This allows the magnetism to vanish at high values of $\rho_i$. The positive parameters $A$ and $B$ are fitted to the specific system.  The model \eqref{EAM_Dudarev} combines aspects from both the Stoner and Ginzburg-Landau models. In \cite{dudarev2005magnetic}, the model \eqref{EAM_Dudarev} was developed for atomic modeling of radiation damage in body-centered cubic (bcc) Fe. The potential accurately describes the stable bcc phase, point-defect behavior, as well as dislocations. In \cite{chiesa2011optimization}, the potential \eqref{EAM_Dudarev} accurately predicts the formation energy of defects, migration barriers, and cluster formation energies. Additionally, it provides good fits for bulk properties such as second-order elastic properties and lattice constants in both magnetic and non-magnetic cases.

Another semi-empirical model based on the non-magnetic semi-empirical model is the expanded magnetic bond-order potential (mBOP), introduced in \cite{PhysRevLett.106.246402}. The mBOP is given by:
\begin{equation} \label{mBOP}
    E = E_{\rm bond}+E_{\rm pair}+\frac{-1}{4}\sum_iI\left(m_i^2-m_{z_i}^2\right).
\end{equation}
The first two terms in \eqref{mBOP} are the same as those in the non-magnetic bond-order potential (BOP) \cite{mrovec2004bond}. These terms account for the attractive bond energy and the pair potential resulting from electrostatic interaction and overlapping repulsion. The last term is derived from the Stoner model and represents the exchange energy. It introduces local exchange interactions, with $I$ being the Stoner exchange integral, which remains the same for a given element. The energy contributions of electron spins depend on the orientation of the local atomic magnetic moments. In \eqref{mBOP}, $m_i$ represents the difference in the numbers of $d$-electrons with spins aligned along the local moment and against it, while $m_{z_i}$ denotes the magnetic moment of the free atom of type $z_i$. Consequently, the favorable state for the system is to be in a ferromagnetic state.

The magnetic BOP was initially developed for bcc Fe \cite{PhysRevLett.106.246402} and has been subsequently applied to the binary Fe-Co alloy system \cite{egorov2023magnetic}. It has been shown that the mBOP can accurately reproduce magnetic states in bcc, face-centered cubic (fcc), hexagonal close-packed (hcp) Fe, as well as amorphous and liquid phases. Additionally, it can distinguish double-layered antiferromagnetic states, ferromagnetic states, and conventional antiferromagnetic states. Surprisingly, the model can even reproduce magnetic moment collapse in compressed systems \cite{friak2001ab, friak2008ab} and vibrational properties of different phases, even though it was not intentionally fitted to any vibrational properties.

In the case of the Fe-Co system, the mBOP was extended by incorporating two additional terms: an embedding energy based on electron density and a repulsion term that includes short-range core repulsion. For the calculation of elastic constants, an environment-dependent Yukawa repulsion term was also included. Overall, the mBOP model offers the flexibility to include various terms to account for specific physical problems, allowing for more qualitative predictions. The accuracy of the mBOP model is reported to reach 10 meV/atom based on comparisons with DFT calculations. The mBOP has been successfully applied in investigating point and planar defects, analyzing the phonon spectrum, determining elastic properties, and even simulating phase transitions \cite{egorov2023magnetic}.

The model, referred to as the efficient interaction model (EIM), was proposed in \cite{li2021_eim, li2022_eim}. This model is based on the combination of non-magnetic and magnetic potentials including generalized form of the Heisenberg model. The EIM can be expressed as:
\begin{equation}
    H = \sum_{i} \sigma_i \cdot \left(A_i \bm{m}_i^2 + B_i \bm{m}_i^4 + \sum_j \sigma_j \cdot J_{ij} {\bm m}_i {\bm m}_j \right) + \sum_{i} \sigma_i \cdot \left(\epsilon_i + \sum_j \sigma_j \cdot (V_{ij} + \alpha_{ij} T) \right),
\end{equation}
where $\sigma_i$ and $\sigma_j$ are occupation variables, equal to 1 (or 0) for occupied (or vacant) lattice sites $i$ and $j$. $A_i$ and $B_i$ represent the on-site magnetic parameters, $J_{ij}$ is the exchange interaction parameter, $\epsilon_i$ is the on-site non-magnetic parameter, and $V_{ij}$ and $\alpha_{ij}$ are non-magnetic interaction parameters. The variable $T$ denotes the absolute temperature. 

In the study presented in \cite{li2022_eim}, the authors investigated phase stability and vacancy formation in fcc Fe-Ni alloys across a broad composition-temperature range using the EIM. The obtained fcc phase diagram using the EIM shows an overall good agreement with the one calculated using the CALPHAD code \cite{saunders1998_calphad}. However, it is worth noting that a limitation of the EIM is its on-lattice nature, which prevents its application in the investigation of materials with complex defects. Additionally, this model does not account for vibrational entropy.

It is pertinent to mention the models developed to describe the specific effects caused by spin-orbit coupling (SOC), with the properties of magnetoelectric multiferroics being a prominent example. In particular, multiferroics include perovskites as $\rm{BiFeO_3}$ and $\rm{(Ho,Tb,Y)MnO_3}$. The models that are capable of reproducing the properties of such materials differs from the ones described above and detailed in \cite{xu2024first}.

The models described above have stood the test of time and continue to serve as reliable frameworks for describing magnetic interactions in solids. As we have seen, these models require the determination or selection of various parameters. Several approaches exist for identifying suitable parameter values, with many involving the combination of experimental data and quantum-mechanical calculations. Quantum-mechanical calculations are typically based on the approximation of solutions to the electronic Schr\"odinger equation. Among the various methods utilized in solid state physics, density functional theory (DFT) stands out as one of the most widely employed. In the following section, we will delve into the principles of DFT.

{\it 3. Density Functional Theory.} Quantum-mechanical models allow for the calculation of properties of materials by solving various approximations of the Schr\"odinger equation:
\begin{equation} \label{Schroedinger}
\hat{H} \Psi = E \Psi,
\end{equation}
where $\Psi$ represents the total wave function of the system, which includes both electronic ${\bf r}_i$ and nuclear ${\bf R}_i$ degrees of freedom. However, in general, solving for the exact solution of \eqref{Schroedinger} is not feasible. 
One commonly used approximation is the Born-Oppenheimer approximation, which decouples the electronic structure from the nuclear motion. This approximation allows for electronic structure calculations under the assumption that the nuclei are fixed in space:

\begin{equation} \label{BO}
\left[\sum_{i=1}^{N_e} \left(-\dfrac{\hbar^2}{2 m_i} \nabla_i^2 \right) + \sum_{i=1}^{N_e}\sum_{j=1}^{N_n} V({\bf r}_i,{\bf R}_j) + \sum_{i=1}^{N_e-1}\sum_{k>i}^{N_e} U({\bf r}_i,{\bf r}_k) \right] \Psi = E \Psi,
\end{equation}
where $E$ represents the total energy, $N_e$ is the number of electrons, $N_n$ is the number of nuclei, $V({\bf r}_i, {\bf R}_j) = \frac{Z_j e^2}{4 \pi \epsilon_0 |{\bf r}_i - {\bf R}_j|}$ denotes the external field generated by the nuclei ${\bf R}_j$, and $U({\bf r}_i,{\bf r}_k)=\frac{e^2}{4 \pi \epsilon_0 |{\bf r}_i - {\bf r}_k|}$ represents the electron-electron interaction energy. However, even with the Born-Oppenheimer approximation, finding an exact solution is still challenging in general cases.

One of the most widely used methods for investigating materials in computational materials science is density functional theory (DFT). DFT was proposed by Hohenberg and Kohn \cite{hohenberg_kohn_1964} and relies on the fact that there is one-to-one correspondence between the ground-state many-electron density and the external potential acting on it. The ground-state energy is then a functional of the ground-state density $n({\bf r})$ (which uniquely determines the ground-state properties of a many-electron system), and the external potential acting on the system. Thus, the total energy $E$ in an external potential $V_{\rm ext.}({\bf r})$ is given by the Hohenberg-Kohn functional:
$$E[n({\bf r})] = F[n({\bf r})] + \int V_{\rm ext.}({\bf r}) n({\bf r}) d^3 {\bf r}.$$
By minimizing this functional with respect to the electron density $n({\bf r})$ we find the ground-state density $n_0$ and the ground-state energy $E_0$. The functional $F$ can be re-written in terms of the kinetic energy of non-interacting electron, the electron-electron interaction energy, and not precisely known electron exchange-correlation term. This results in the Kohn-Sham functional:
\begin{equation} \label{KSfunctional}
E = E(\{\psi_i\})= \sum_{i=1}^{N_e} \int \psi_i^* \left(-\dfrac{\hbar^2}{2 m_i} \nabla_i^2 \right) \psi_i d^3 {\bf r} + \dfrac{1}{2} \dfrac{e^2}{4 \pi \epsilon_0} \int \dfrac{n({\bf r}) n({\bf r}')}{|{\bf r}-{\bf r}'|} d^3 {\bf r} d^3 {\bf r}'+E_{XC}[n({\bf r})] + \int V_{\rm ext.}({\bf r}) n({\bf r}) d^3 {\bf r},
\end{equation}
where the electron density 
\begin{equation} \label{electron_density}
n({\bf r}) = \sum \limits_{i=1}^{N_e} \int \psi_i^* ({\bf r}) \psi_i ({\bf r}) d^3 {\bf r} 
\end{equation}
depends on the electron wavefunctions $\psi_i ({\bf r})$. For finding $\psi_i ({\bf r})$ minimizing the functional \eqref{KSfunctional}, we formulate the so-called Kohn-Sham (KS) equations \cite{kohn_sham_1965}:
\begin{equation} \label{KSequations}
\left( -\dfrac{\hbar^2}{2 m_i} \nabla_i^2 + \dfrac{e^2}{4 \pi \epsilon_0} \int \dfrac{n({\bf r}')}{|{\bf r}-{\bf r}'|} d^3 {\bf r}' + V_{XC}[n({\bf r})] +  V_{\rm ext.}({\bf r}) \right) \psi_i({\bf r}) = \left( -\dfrac{\hbar^2}{2 m_i} \nabla_i^2 + V_{\rm eff.}({\bf r}) \right) \psi_i({\bf r}) = {\lambda_i} \psi_i({\bf r}),
\end{equation}
where $V_{XC}[n({\bf r})] = \dfrac{\delta E_{XC}[n({\bf r})]}{\delta n({\bf r})}$, $V_{\rm eff.}({\bf r})$ is the effective potential, and $\lambda_i$, $\psi_i$ are the solutions of KS equations. Thus, instead of solving many-particle problem \eqref{BO}, we solve independent single-particle KS equations \eqref{KSequations}. These equations need to be solved self-consistently: we start from any randomly initialized density $n({\bf r})$, next we calculate the effective potential $V_{\rm eff.}({\bf r})$, after that we find $\psi_i({\bf r})$ from \eqref{KSequations} on the next iteration, update $n({\bf r})$ using \eqref{electron_density}, and compare it to the one from the previous iteration. We continue this process until the convergence.

In DFT, the exact form of the exchange-correlation potential $V_{XC}$ in the Schr\"odinger equation \eqref{Schroedinger} is not known and needs to be approximated. One commonly used approximation is the local density approximation (LDA), where it is assumed that the exchange-correlation energy per electron at a given point is equal to that of a homogeneous electron gas with the same density. Additionally, it is assumed that the exchange-correlation energy functional is purely local, depending solely on the electron density $n({\bf r})$ \cite{kohn_sham_1965}.

 To achieve a higher accuracy, the generalized gradient approximation (GGA) is employed. In GGA, the exchange-correlation energy depends not only on $n({\bf r})$ but also on its gradient $\nabla n({\bf r})$ \cite{becke1988_gga, perdew1996_gga}. This approach captures the non-local behavior of the electron density.

 More accurate yet computationally expensive approximations to $V_{XC}$ include the meta-GGA, which incorporates the kinetic energy density $\nabla^2 n({\bf r})$ and was proposed in \cite{perdew1999_meta_gga}. Among the popular meta-GGA functionals is the strongly constrained and appropriately normed (SCAN) functional \cite{sun2015_SCAN}. 
 
 Hybrid functionals represent another class of approximations for $V_{XC}$ \cite{becke1993_hybrid}. These functionals combine a fraction of the exact Hartree-Fock exchange with the exchange-correlation energy from LDA or GGA functionals. Prominent examples of hybrid functionals include B3LYP \cite{stephens1994_B3LYP}, PBE0 \cite{perdew1996_PBE0}, PBEsol \cite{constantin2009_PBEsol}, HSE06 \cite{krukau2006_HSE06}, and HSEsol \cite{schimka2011_HSEsol}.

The DFT method can be extended to spin-polarized calculations by introducing spins $s_j$, $j=1,\ldots,N_e$ in the $N_e$-electron wavefunction: $\psi_i = \psi_i({\bf r}_1, s_1, \ldots, {\bf r}_{N_e}, s_{N_e})$. In this context, spins can take the values $+\dfrac{1}{2}$ and $-\dfrac{1}{2}$. Consequently, the electron density is given by:
\begin{equation} \label{electron_density_up_down}
\begin{array}{c}
\displaystyle
n({\bf r}) = \sum \limits_{i=1}^{N_e} \Bigl( \int d s_2 d^3 {\bf r}_2 \ldots \int d s_{N_e} d^3 {\bf r}_{N_e} \psi_i^* \left({\bf r}, +\dfrac{1}{2}, {\bf r}_2, s_2, \ldots, {\bf r}_{N_e}, s_{N_e}\right) \psi_i \left({\bf r}, +\dfrac{1}{2}, {\bf r}_2, s_2, \ldots, {\bf r}_{N_e}, s_{N_e}\right) +
\\
\displaystyle
\int d s_2 d^3 {\bf r}_2 \ldots \int d s_{N_e} d^3 {\bf r}_{N_e} \psi_i^* \left({\bf r}, -\dfrac{1}{2}, {\bf r}_2, s_2, \ldots, {\bf r}_{N_e}, s_{N_e}\right) \psi_i \left({\bf r}, -\dfrac{1}{2}, {\bf r}_2, s_2, \ldots, {\bf r}_{N_e}, s_{N_e} \right) \Bigr) = 
\\
\displaystyle
\sum \limits_{i=1}^{N_e} \int \psi_i^{\uparrow *} ({\bf r}) \psi_i^{\uparrow} ({\bf r}) d^3 {\bf r} + \sum \limits_{i=1}^{N_e} \int \psi_i^{\downarrow *} ({\bf r}) \psi_i^{\downarrow} ({\bf r}) d^3 {\bf r} = n^{\uparrow}({\bf r}) + n^{\downarrow}({\bf r}),
\end{array}
\end{equation}
where $\psi_i^{\uparrow} ({\bf r})$, $n^{\uparrow}({\bf r})$ correspond to ``spin-up'' electrons and $\psi_i^{\downarrow} ({\bf r})$, $n^{\downarrow}({\bf r})$ correspond to ``spin-down'' electrons. Additionally, along with the electron density, we define the spin density:
\begin{equation} \label{spin_density}
\begin{array}{c}
m({\bf r}) = n^{\uparrow}({\bf r}) - n^{\downarrow}({\bf r}).
\end{array}
\end{equation}
In spin-polarized DFT calculations, the Kohn-Sham (KS) equations has the following form:
\begin{equation} \label{KSequations_up_down}
\begin{array}{c}
\displaystyle
\left( -\dfrac{\hbar^2}{2 m_i} \nabla_i^2 + \dfrac{e^2}{4 \pi \epsilon_0} \int \dfrac{n({\bf r}')}{|{\bf r}-{\bf r}'|} d^3 {\bf r}' + V_{XC}[n^{\uparrow}({\bf r}),n^{\downarrow}({\bf r})] +  V_{\rm ext.}({\bf r}) \right) \psi_i^\uparrow({\bf r}) = \lambda_i^{\uparrow} \psi_i^{\uparrow}({\bf r}),
\\
\displaystyle
\left( -\dfrac{\hbar^2}{2 m_i} \nabla_i^2 + \dfrac{e^2}{4 \pi \epsilon_0} \int \dfrac{n({\bf r}')}{|{\bf r}-{\bf r}'|} d^3 {\bf r}' + V_{XC}[n^{\uparrow}({\bf r}),n^{\downarrow}({\bf r})] +  V_{\rm ext.}({\bf r}) \right) \psi_i^\downarrow({\bf r}) = \lambda_i^{\downarrow} \psi_i^{\downarrow}({\bf r}),
\end{array}
\end{equation}
where $\lambda_i^{\uparrow}$, $\psi_i^{\uparrow}$ and $\lambda_i^{\downarrow}$, $\psi_i^{\downarrow}$ are the solutions. Here the KS equations for $\psi_i^\uparrow({\bf r})$ and $\psi_i^\downarrow({\bf r})$ are coupled due to the electron-electron, exchange-correlation, and external potentials. The specific forms of the exchange-correlation functional associated with the ``spin-up'' and ``spin-down'' densities follow the approaches mentioned earlier, such as the local spin density approximation (LSDA) \cite{parr_yang_1993_dft}, which is based on the local density approximation (LDA).

Here we described the so-called non-relativistic DFT. For the principles of relativistic DFT, refer, e.g., to \cite{jacob2012_spin_in_DFT} and related literature. 
Additionally, it is worth noting that the magnetic moment of the $j$-th atom can be determined by integrating the spin density \eqref{spin_density} over a specific region $\Omega_j$ around the atom, depending on the chosen partitioning scheme:

\begin{equation} \label{magnetic_moment}
    m_j = \int \limits_{\Omega_j} m({\bf r}) d {\bf r}.
\end{equation}
This means that magnetic moments can be treated as an additional degree of freedom along with atomic positions when using magnetic machine-learning interatomic potentials. In the following section, we discuss the results obtained in the investigation of magnetic materials using magnetic machine-learning potentials.

The DFT method has been implemented in various software packages, such as VASP \cite{VASP1,VASP2,VASP3,VASP4}, ABINIT \cite{ABINIT_2002}, and Quantum Espresso \cite{QuantumEspresso_2009}, which are commonly used for the calculation of solids. DFT has proven to be successful in studying magnetic materials, with notable achievements in the case of Fe, one of the most abundant magnetic materials on Earth. These accomplishments include accurate predictions of phonon spectra at different temperatures \cite{kormann2012-fe-phonon,kormann2014-magnon-phonon}, phase transition from bcc lattice to fcc lattice (bcc-fcc phase transition) \cite{Leonov2012-fe-bcc-fcc}, and self-diffusion coefficient at different temperatures \cite{versteylen2017-fe-diffusion}. Furthermore, DFT calculations have been employed for investigating the magnetic and thermodynamic properties of FeCr alloys \cite{klaver2006-fe-cr}, studying FeCo alloys \cite{diaz2006-fe-co}, and exploring the structural stability and phase diagram of CrN \cite{zhou2014-cr-n}. Additionally, DFT has proven powerful for investigating more complex magnetic alloys, involving up to five components. A comprehensive review of simulating magnetic alloys with DFT can be found in \cite{ikeda2019_DFT_HEA}.

Despite the considerable success of DFT in investigating various magnetic materials, its computational cost remains a significant drawback. The scaling of Kohn-Sham (KS) DFT with the number of electrons, denoted as $N_e$, follows an $O(N_e^3)$ complexity. Consequently, DFT is most effectively applied to systems containing up to several hundred atoms. However, this number of atoms may be insufficient for studying processes such as diffusion, phase transitions, or atomic layer deposition. To address this challenge, machine-learning interatomic potentials (MLIPs) have emerged as an alternative approach that combines the efficiency of empirical potentials with the accuracy of DFT calculations. MLIPs offer a promising solution in bridging the gap between accuracy and computational efficiency.

\begin{figure}[h!]
    \centering
    \includegraphics[width=\textwidth]{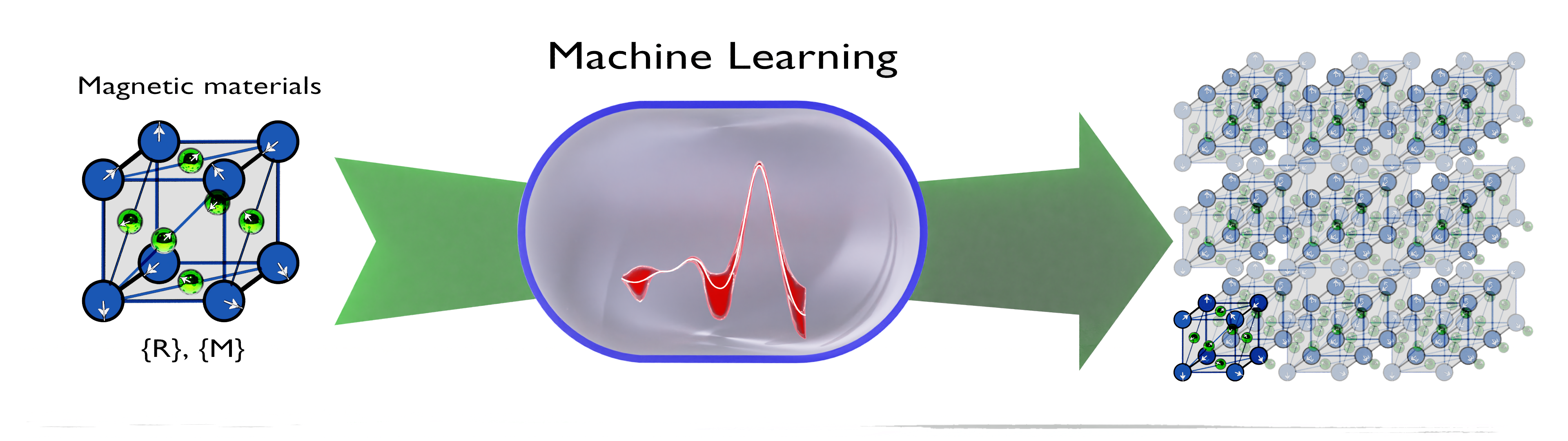}
    \caption{Illustration of the transformative impact of machine-learning interatomic potentials (MLIPs) on computational materials science. Density Functional Theory (DFT) calculations allows investigating the systems up to several hundred of atoms and can be used for creating a training set for machine-learning interatomic potentials (MLIPs). These MLIPs, in turn, enable the exploration of larger atomic systems, exceeding the time and length scales limitations of DFT calculations. Consequently, MLIPs have emerged as a reliable and powerful tool, expanding the possibilities of materials research.}
    \label{fig:dft_ml}
\end{figure}

{\it 4. Machine-learning interatomic potentials.} Over the past 17 years, MLIPs have established themselves as a reliable tool for the theoretical study of materials. MLIPs offer the ability to extend the time and length scales of DFT calculations. We illustrate the latter in Fig.~\ref{fig:dft_ml}. These potentials incorporate coefficients, also known as parameters, which can be determined based on DFT data through a process called training or fitting. It is worth noting that empirical potentials can also be fitted to DFT data. However, MLIPs distinguish themselves by allowing for an increase in the number of parameters, which enables reaching the accuracy of DFT calculations. In other words, MLIPs are systematically improvable interatomic potentials and this stands as a key difference between MLIPs and empirical potentials.

Initially, MLIPs were developed for the investigation of non-magnetic materials. Consequently, they were limited to studying only one magnetic, such as nonmagnetic, or ferromagnetic, or antiferromagnetic. These potentials approximate the potential energy as a sum of atomic energies dependent on the local environments. The functional forms of these potentials typically explicitly include atomic positions, atomic types, and lattice parameters of configurations. Prominent MLIP models include the high-dimensional neural network potential (HDNNP) \cite{behler_generalized_2007}, Gaussian approximation potential (GAP) \cite{bartok_gaussian_2010}, spectral neighbor analysis potential (SNAP) \cite{thompson_spectral_2015}, and moment tensor potential (MTP) \cite{Shapeev2016-MTP}. More recent MLIP developments include the atomic cluster expansion (ACE) \cite{drautz_ace_2019}, DeepMD \cite{wang2018-deepmd}, physically informed artificial neural network (PINN) \cite{pun2019-pinn}, and neural equivariant interatomic potentials (NequIP) \cite{batzner2022-equivarnn}.

The aforementioned MLIPs do not explicitly incorporate magnetic degrees of freedom, such as magnetic moments, into their functional forms, making them non-magnetic by default. This limitation is not critical when considering a single magnetic state or a small set of states, as demonstrated in studies like \cite{dragoni2018_gap_fm_fe}, where a non-magnetic GAP was used to investigate ferromagnetic bcc Fe. However, for studying magnetic materials at finite temperatures, a single magnetic state description becomes insufficient due to the thermal fluctuations of atomic magnetic moments and spin flips. To properly describe a range of paramagnetic states and predict Curie and Neel temperatures, it is crucial to explicitly include magnetic moments in the functional forms of MLIPs. In recent years, efforts have been made to extend some of the aforementioned MLIPs to incorporate different magnetic states. We discuss the magnetic MLIPs in this section.

To construct a magnetic MLIP, one approach is to combine a non-magnetic potential with the (generalized) Heisenberg model. In the work \cite{nikolov2021_snap_heisenberg}, the authors introduced a hybrid potential by combining the spectral neighbor analysis potential (SNAP) with the Heisenberg model. The resulting form is as follows:
\begin{equation}
    H_{\rm mag} = - \sum_{i \neq j} J(\bm{r}_{ij}) ({\bm m}_i \cdot {\bm m}_j - 1) - \sum_{i \neq j} K(\bm{r}_{ij}) (({\bm m}_i \cdot {\bm m}_j)^2 - 1),
\end{equation}
where ${\bm m}_i$ and ${\bm m}_j$ are magnetic moments located on atoms $i$ and $j$. The magnetic exchange functions, $J(\bm{r}_{ij})$ and $K(\bm{r}_{ij})$ (in eV), represent the interaction between these moments, where $\bm{r}_{ij}$ represents the distance between atoms $i$ and $j$. The investigation focused on the systems of bcc, fcc, and hcp Fe. The authors parameterized the Heisenberg model by fitting $J(\bm{r}_{ij})$ and $K(\bm{r}_{ij})$ to spin-spiral DFT calculations. Subsequently, they constructed a training set using spin-polarized DFT calculations and subtracted the magnetic energy, forces, and stresses contributions calculated with the Heisenberg model from each atomic configuration in the training set. Finally, the authors fitted a non-magnetic version of the SNAP potential to the DFT data, excluding the magnetic contributions. By combining the SNAP and Heisenberg model, the authors were able to calculate various properties and characteristics of Fe, such as elastic constants, bulk modulus, lattice constants, magnetization, and specific heat. Notably, the calculated Curie temperature exhibited good agreement with experimental observations. To achieve this agreement, the authors employed pressure- and magnetization-controlled calculations in combination with the temperature-rescaling method \cite{evans2015quantitative} in their spin dynamics calculations. All calculations were performed using a $20 \times 20 \times 20$ bcc supercell.

In the work by Chapman et al. \cite{chapman2022_nn_defect_Fe}, the authors combined the Embedded Atom Method (EAM) to account for non-magnetic interactions, the Heisenberg-Landau model to describe magnetic interactions, and a neural network incorporating magnetic moments to capture interactions not included in EAM and the Heisenberg-Landau model. This combined model was trained on an extensive dataset comprising over 10000 configurations derived from DFT simulations. 

The fitted combined model was then employed to investigate defective Fe. Notably, the model successfully reproduced the cohesive energy of bcc and fcc Fe in various magnetic states. Moreover, it accurately predicted the formation energy and intricate magnetic structures of point defects, demonstrating quantitative agreement with DFT. The model also captured the magnetic behavior near the core of defects, including the reversal and quenching of magnetic moments. Furthermore, the predicted Curie temperature exhibited good agreement with experimental observations.
To calculate the Curie temperature, the authors considered a $20 \times 20 \times 20$ supercell accommodating a total of 16000 atoms.

In the paper by Eckhoff et al. \cite{eckhoff2021_mhdnnp}, the authors introduced a different approach to constructing a magnetic MLIP. They extended the high-dimensional neural network potential (HDNNP) to account for magnetic moments and created the magnetic HDNNP (mHDNNP). In mHDNNP, the atomic energy contribution of the $i$-th atom is represented by the following expression:

\begin{equation} \label{HDNNP_SE}
    E_i = f_a^2 \left[w_{01}^2 + \sum \limits_{k=1}^3 w_{k1}^2 f_a^1 \left(w_{0k}^1 + \sum \limits_{\mu=1}^2 w_{\mu k}^1 G_i^{\mu} \right) \right],
\end{equation}
where $w_{kl}^m$ represents the weight parameter connecting node $l$ in layer $m$ with node $k$ in layer $m-1$. The term $f_a^m$ denotes the activation function used in the neural network. The two-body interactions (radial part) are described by $G_i^{1}$, while the three-body interactions (angular part) are captured by $G_i^{2}$. It is important to note that mHDNNP only considers collinear spin interactions.
To represent the magnetic configuration, the authors introduced an atomic spin coordinate:
\begin{equation} \label{spin_coordinate}
s_i = \begin{cases}
0, |M_S| < M_S^{\rm thres}\\
1, {\rm otherwise}.
\end{cases}
\end{equation}
In this equation, $M_S = \dfrac{1}{2} (n^{\uparrow} - n^{\downarrow})$, where $n^{\uparrow}$ and $n^{\downarrow}$ represent the ``up'' and ``down'' spin populations, respectively. The parameter $M_S^{\rm thres}$ is introduced as a threshold value to filter out noise in the spin reference data, which arises from the ambiguity in assigning spins in electronic structure calculations.
The radial part describing two-body interactions has the following form:
\begin{equation}
    G_i^1 = \sum_j M^x(s_i, s_j) \cdot e^{-\eta \bm{r}_{ij}^2} \cdot f_c(|\bm{r}_{ij}|),
\end{equation}
where $\eta$ represents a parameter of the potential, while $M^x$ corresponds to the radial spin-augmentation functions (SAFs), $x=0,+,-$, $M^0(s_i,s_j)=1$, $M^+(s_i,s_j) = \dfrac{1}{2}|s_i s_j|\cdot|s_i+s_j|$, $M^-(s_i,s_j) = \dfrac{1}{2}|s_i s_j|\cdot|s_i-s_j|$, $f_c$ represents the cut-off function, and $|\bm{r}_{ij}|$ denotes the distance between atoms. The radial spin-augmentation functions can be categorized as follows: $M^0$ describes interactions between two non-magnetic atoms or between a non-magnetic and a magnetic atom, $M^+$ represents interactions between atoms with parallel spins (ferromagnetic interactions), and $M^-$ represents interactions between atoms with antiparallel spins (antiferromagnetic interactions). Similarly, an angular part is incorporated in the following manner:
\begin{equation}
G_i^2 = 2^{- \zeta} \sum_j \sum_{k \neq j} M^{xx} (s_i,s_j,s_k) \cdot [1 + \lambda {\rm cos}(\theta_{ijk})]^{\zeta} \cdot e^{-\eta \left(\bm{r}_{ij}^2 + \bm{r}_{jk}^2 + \bm{r}_{ik}^2 \right)} \cdot f_c(|\bm{r}_{ij}|) \cdot f_c(|\bm{r}_{jk}|) \cdot f_c(|\bm{r}_{ik}|).
\end{equation}
In this context, $\zeta$ and $\lambda$ represent parameters of the potential, while $\theta_{ijk}$ denotes the angle $j-i-k$ involving the neighboring atoms $j$ and $k$. $M^{xx}$ stands for the angular SAFs, where $xx$ can take values of $00,++,--,$ or $+-$. These SAFs allow differentiation between three scenarios where $s_i \neq 0$, $s_j \neq 0$, and $s_k \neq 0$: 1) $s_i = s_j = s_k$, 2) $s_i \neq s_j = s_k$, and 3) $s_i = s_j \neq s_k$ (or $s_i = s_k \neq s_j$). Specifically, $M^{++}$, $M^{--}$, and $M^{+-}$ evaluate to one and describe scenarios 1), 2), and 3) respectively (refer to \cite{eckhoff2021_mhdnnp} for further details). $M^{00}$ is set to one and characterizes interactions involving more than one $s = 0$ or $s_i = 0$. Likewise, $M^{++}$ and $M^{--}$ distinguish ferromagnetic and antiferromagnetic interactions, akin to the radial $M^+$ and $M^-$ SAFs.

The mHDNNP approach was applied by the authors to investigate the Mn-O system. They utilized mHDNNP in molecular dynamics simulations, including Monte Carlo spin flips (MDMC), for a Mn-O supercell of size $6 \times 6 \times 6$ (containing approximately 1700 atoms). The Neel temperature was calculated, and the linear thermal expansion coefficient, which exhibited good agreement with experimental observations, was explored. Additionally, the authors investigated the temperature dependence of magnetization, and they observed that an increasing Mn vacancy concentration resulted in a decrease in the Neel temperature.

Another magnetic MLIP model with collinear spins was proposed in \cite{Novikov2022-mMTP} for single-component materials and generalized to multi-component materials in \cite{Kotykhov2023-cDFT-mMTP}. In these papers, the authors introduced the magnetic Moment Tensor Potential (mMTP). They extended the original Moment Tensor Potential (MTP) \cite{Shapeev2016-MTP} by incorporating the collinear magnetic moments of the atoms into the functional form. Notably, mMTP is formulated as a polynomial potential. The atomic energy contribution can be expressed as follows:
\begin{equation} \label{MTP_site_energy}
    E(\mathfrak{n}_i) = \sum_{\alpha} \xi_{\alpha} B_{\alpha}(\mathfrak{n}_i),
\end{equation}
where $\mathfrak{n}_i$ represents the atomic neighborhood of the $i$-th atom, $\xi_{\alpha}$ corresponds to the parameters to be optimized, and $B_{\alpha}$ denotes a basis function defined using the moment tensor descriptors as follows: 
\begin{equation} \label{MTPdescriptors}
M_{\mu,\nu}(\mathfrak{n}_i)=\sum_{j} f_{\mu}(|\bm{r}_{ij}|,z_i,z_j,m_i,m_j) \underbrace {\bm{r}_{ij}\otimes...\otimes \bm{r}_{ij}}_\text{$\nu$ times},
\end{equation}
where $``\otimes"$  represents the outer product of vectors, thus making the angular part $\bm{r}_{ij}\otimes...\otimes \bm{r}_{ij}$ a tensor of rank $\nu$. The function $f_{\mu}(|\bm{r}_{ij}|,z_i,z_j,m_i,m_j)$ is a polynomial that depends on the distance $|\bm{r}_{ij}|$ between atoms, the atomic types $z_i$ and $z_j$, and the magnetic moments $m_i$ and $m_j$; it can be expressed as:
\begin{equation}\label{MTPRadialFunction}
f_{\mu}(|\bm{r}_{ij}|,z_i,z_j,m_i,m_j) = \sum_{\zeta} \sum_{\gamma} \sum_{\beta} c^{\beta, \gamma, \zeta}_{\mu,z_i,z_j} \psi_{\beta}(m_i) \psi_{\gamma}(m_j) 
\varphi_\zeta (|\bm{r}_{ij}|) (r_{\rm cut} - |\bm{r}_{ij}|)^2,
\end{equation}
where $c^{\beta, \gamma, \zeta}_{\mu,z_i,z_j}$ represent the ``radial'' parameters to be optimized. The function $\varphi_{\zeta} (|\bm{r}_{ij}|)$ corresponds to a polynomial basis function defined on the interval $(r_{\rm min}, r_{\rm cut})$, where $r_{\rm min}$ denotes the minimal distance between atoms and $r_{\rm cut}$ stands for the cut-off radius beyond which atoms do not interact. The other functions, $\psi_{\beta}(m_i)$ and $\psi_{\gamma}(m_j)$, represent the polynomial basis functions of the local magnetic moments defined on the interval $(-M_{\rm max}^{z_i}, M_{\rm max}^{z_i})$, with $M_{\rm max}^{z_i}$ being the maximal absolute value of the magnetic moment for the atomic type $z_i$ in a training set. We note that in \eqref{MTPRadialFunction}, the magnetic moments are determined using \eqref{magnetic_moment}.

The basis functions $B_{\alpha}$ in mMTP are defined as all possible contractions of $M_{\mu,\nu}(\mathfrak{n}_i)$ to a scalar, such as:
\[
M_{1,0}(\mathfrak{n}_i), ~M_{0,1}(\mathfrak{n}_i) \cdot M_{1,1}(\mathfrak{n}_i),  ~M_{3,2}(\mathfrak{n}_i):M_{1,2}(\mathfrak{n}_i), ~\ldots\,,
\]
where $``\cdot"$ represents the dot product of two vectors, while $``:"$ denotes the Frobenius product of two matrices. In theory, an infinite number of such mMTP basis functions could be generated, but in practice, this number is constrained by the so-called level of MTP (refer to \cite{Novikov2022-mMTP} for more information). During an atomistic simulation, mMTP operates in two steps: first, magnetic moments are equilibrated at fixed atomic positions and lattice parameters, and then the atomic positions and lattice parameters change while keeping the magnetic moments fixed.

In \cite{Novikov2022-mMTP}, the authors applied mMTP to investigate the system of Fe and demonstrated the capability to reproduce various properties of Fe. The fitted mMTP, trained on more than 9000 structures generated with DFT, successfully reproduced the phonon spectra in paramagnetic and ferromagnetic states, energy/volume curves, and the dependence of absolute equilibrium magnetic moments on the volume of bcc, fcc, and hcp Fe in both ferromagnetic and antiferromagnetic states. Furthermore, the mMTP accurately predicted DFT energies and local magnetic moments at finite temperature.

In a more recent study \cite{Kotykhov2023-cDFT-mMTP}, mMTP was employed to investigate the Fe-Al alloy system with varying concentrations of Al and Fe. The authors demonstrated that mMTP calculations yielded formation energies, equilibrium lattice parameters, and total magnetic moments of the unit cell for various Fe-Al structures that were in good agreement with those obtained using DFT. Unlike in \cite{Novikov2022-mMTP}, the training set for the Fe-Al system was constructed using constrained DFT (cDFT) calculations \cite{Gonze_2022}, which allowed for calculating configurations with excited (non-equilibrium) magnetic moments. By fitting mMTP to this training set, the equilibration of magnetic moments in a configuration was achieved, obviating the need for an extensive training set as in \cite{Novikov2022-mMTP}. The training set for Fe-Al consisted of approximately 2000 configurations, providing accurate predictions of the aforementioned properties for Fe-Al.

A limitation of mHDNNP and mMTP is that they consider only collinear magnetic moments within their functional forms. Recently, several MLIPs have been developed to incorporate non-collinear magnetic moments. In \cite{drautz2024_noncolACE}, the authors introduced the concept of non-collinear magnetic Atomic Cluster Expansion (mACE). They defined state variables $\sigma_{ji} = (z_j, \bm{r}_{ji}, \bm{m}_j)$ for atom $j$ neighboring atom $i$, which account for interatomic distance vectors $\bm{r}_{ji}$, chemical species $z_j$, and magnetic moments $\bm{m}_j$. Then, the authors introduced single bond basis functions as follows:
\begin{equation}\label{ACE_phi}
\phi_{z_j z_i nlmn'l'm'}(\sigma_{ji}) = R_{nl}^{z_j z_i} (|\bm{r}_{ji}|) Y_l^m (\hat{\bm{r}}_{ji}) M_{n'l'}^{z_j z_i} (|\bm{m}_j|) Y_{l'}^{m'} (\hat{\bm{m}}_{j}),
\end{equation}
where $\hat{\bm{r}}_{ji} = \bm{r}_{ji}/|\bm{r}_{ji}|$, $\hat{\bm{m}}_{j} = \bm{m}_j/|\bm{m}_j|$, $Y_l^m$ and $Y_{l'}^{m'}$ are spherical harmonics, $R_{nl}^{z_j z_i}$ and $M_{n'l'}^{z_j z_i}$ are radial functions related to the distances between atoms and the magnetic moments of atoms. The single bond atomic basis function:
\begin{equation}\label{ACE_basis_single}
A_{iz_i z_j nlmn'l'm'} = \sum \limits_j \delta_{z z_j} \phi_{z_i z_j nlmn'l'm'}(\sigma_{ji}).
\end{equation}
Permutation-invariant many-body basis functions are generated by taking products as follows:
\begin{equation}\label{ACE_basis}
\bm{A}_{i \bm{z} \bm{n} \bm{l} \bm{m} \bm{n}' \bm{l}' \bm{m}'} = \prod_{t=1}^{\nu} A_{i z_i^t z_j^t n^t l^t m^t n^{'t} l^{'t} m^{'t} }.
\end{equation}
The products follow a body order of $\nu+1$. The vectors $z$, $n$, $l$, $m$, $n'$, $l'$, and $m'$ have a length of $\nu$. In the case of non-collinear magnetic ACE, the energy $E_i$ of atom $i$ can be expressed as follows:
\begin{equation}\label{ACE_site_energy}
E_i = \sum \limits_{ \bm{z} \bm{n} \bm{l} \bm{m} \bm{n}' \bm{l}' \bm{m}'} \tilde{\bm{c}}_{z_i \bm{z} \bm{n} \bm{l} \bm{m} \bm{n}' \bm{l}' \bm{m}'} \bm{A}_{i \bm{z} \bm{n} \bm{l} \bm{m} \bm{n}' \bm{l}' \bm{m}'},
\end{equation}
where $\tilde{\bm{c}}_{z_i \bm{z} \bm{n} \bm{l} \bm{m} \bm{n}' \bm{l}' \bm{m}'}$ are the parameters of non-collinear magnetic ACE.

In the work \cite{drautz2024_noncolACE}, non-collinear magnetic ACE was tested using the prototypical magnetic element Fe. ACE was fitted to DFT data and successfully reproduced many different properties and quantities of Fe predicted by DFT across different phases and magnetic states. These included energy/volume curves, the relationship between magnetic moment and unit cell volume, elastic constants, phonon spectra, and defect formation energies. The authors also predicted the Curie temperature and temperatures of phase transitions from bcc to fcc and from fcc to bcc using ACE. To achieve this, they conducted MDMC simulations involving 3456 atoms. The ACE Curie temperature was underestimated compared to experimental values due to the neglect of lattice thermal expansion. The temperature of the fcc to bcc phase transition was accurately predicted. However, the temperature of the bcc to fcc phase transition was significantly overestimated (by 250 K) due to insufficient consideration of the effect of magnetic fluctuations on the free energy difference. Notably, one of the most challenging aspects of \cite{drautz2024_noncolACE} was the creation of a training set consisting of 70000 structures.

Another non-collinear magnetic MLIP was developed in \cite{domina2022_spinSNAP}. The authors extended the SNAP framework to accommodate arbitrary vectorial fields. To achieve this, they associated a vector $\bm{v}_i$ with each atomic position $\bm{r}_i$ of the $i$-th atom, which can, for instance, represent the local magnetic moment. The vector $\bm{v}_i$ is defined as:
\begin{equation}
    \bm{v}_i = \sum \limits_{q=0, \pm 1} v_{i,q} \hat{\bm{e}}_q ~\rm{with}~
\begin{cases}
v_{i,\pm 1} = \mp \dfrac{1}{\sqrt{2}} (v_{i,x} \mp i v_{i,y}),\\
v_{i,0} = v_{i,z},
\end{cases}
\end{equation}
$\hat{\bm{e}}_{\pm 1} = \mp \dfrac{1}{\sqrt{2}} (\hat{\bm{e}}_x \pm i \hat{\bm{e}}_y)$, $\hat{\bm{e}}_0 = \hat{\bm{e}}_z$, $\hat{\bm{e}}_x$, $\hat{\bm{e}}_y$, $\hat{\bm{e}}_z$ are the unit vectors along $x$, $y$, $z$. Using the defined vector $\bm{v}_a$ the authors further introduced the local vector density:
\begin{equation} \label{SNAP_density}
    \bm{\rho}(\bm{r}) = \sum \limits_{n = 0}^{n_{\rm{max}}} \sum \limits_{l = 0}^{n} \sum \limits_{m = -l}^{l} \sum \limits_{q = 0,\pm 1} c_{nlmq} R_{nl} (r) Y_l^m (\hat{\bm{r}}) \hat{\bm{e}}_q,
\end{equation}
where $n_{\rm max}$ defines a size of the radial basis $R_{nl} (r)$ depending on the distance $r$ between atoms, $Y_l^m (\hat{\bm{r}})$ are the three-dimensional spherical harmonics, and the coefficients $c_{nlmq}$ are calculated as:
\begin{equation}
    c_{nlmq} = \sum \limits_i z_i v_{i,q} \int dr d \hat{\bm{r}} r^2 R_{nl} (r) Y_l^{m*} (\hat{\bm{r}}) \delta (\bm{r} - \bm{r}_i).
\end{equation}
Local vector density \eqref{SNAP_density} in Dirac notation can be written as:
\begin{equation}
    |\bm{\rho} \rangle = \sum \limits_{nlmq} c_{nlmq} |nlmq \rangle,
\end{equation}
where $\langle \bm{r} | nlm \rangle = R_{nl} (r) Y_l^m (\hat{\bm{r}})$ and $q \equiv \hat{\bm{e}}_q$. The authors subsequently formulated all the pertinent quantities using the basis $|nl1JM \rangle$, wherein the combined angular momenta are represented as $\bm{L+1=J}$:
\begin{equation} \label{basis_nlJM}
    |nl1JM \rangle = \sum \limits_{m=-l}^l \sum \limits_{q=-1}^1 C_{lm1q}^{JM} |nlmq \rangle,
\end{equation}
where Clebsch-Gordan coefficients, denoted as $C_{lm1q}^{JM}$, are utilized in this context, the subscript ``1'' represents the angular momentum of $\bm{1}$. The quantum numbers $J$ and $M$ correspond to the total angular momentum and its projection, respectively. Notably, the values of $J$ and $M$ satisfy the conditions $|l-1| \leq J \leq l+1$ and $-J \leq M \leq J$. Through the inversion of \eqref{basis_nlJM}, the authors expressed $|nlmq \rangle$ in terms of $|nlJM \rangle$:
\begin{equation}
    |nlmq \rangle = \sum \limits_{JM} C_{lm1q}^{JM} |nlJM \rangle
\end{equation}
and re-wrote the local vector density as:
\begin{equation}
    |\bm{\rho} \rangle = \sum \limits_{nlJM} u_{nlJM} |nlJM \rangle ~{\rm{with}}~ u_{nlJM} = \langle nlJM | \bm{\rho} \rangle = \sum \limits_{mq} C_{lm1q}^{JM} \langle nlmq | \bm{\rho} \rangle = \sum \limits_{mq} C_{lm1q}^{JM} c_{nlmq}.
\end{equation}
To construct a functional form of spin SNAP, the authors introduced the concept of the power spectrum defined as:
\begin{equation}
    p_{nlJ}^{(i)} = \sum \limits_{M} |u_{nlJM}^{(i)}|^2
\end{equation}
for each atom indexed by $i$. The energy was then expressed as a linear combination of power spectrum vectors using the following expression:
\begin{equation} \label{spinSNAP}
    E = \sum \limits_{nlJ} \theta_{nlJ} \sum \limits_i p_{nlJ}^{(i)},
\end{equation}
where $\theta_{nlJ}$ represent the parameters of spin SNAP. In the paper \cite{domina2022_spinSNAP}, the authors tested spin SNAP using the system of bcc Fe. To create training sets, the authors utilized the Heisenberg model (which excludes longitudinal spin excitations) and the Heisenberg-Landau model (which includes longitudinal spin excitations). This approach was chosen to bypass the time-consuming nature of DFT calculations, as the primary goal was to test spin SNAP. The results demonstrated that spin SNAP, based on the power spectrum, was capable of accurately describing the entire energy surface by effectively extrapolating beyond the energy range covered by the training set, even without the inclusion of longitudinal spin excitations. However, when fitting the model on a training set that incorporated longitudinal spin excitations, spin SNAP accurately reproduced the energies of the configurations included in the training set. Nevertheless, the model's performance was less satisfactory for configurations beyond the energy range covered by the training set. To address this limitation, a non-linear model could be employed.

Several other non-collinear magnetic MLIPs are based on neural networks. In \cite{yu2022_spinGNN}, the authors introduced the spin-dependent graph neural network (SpinGNN) for magnetic materials. The training set consisted of around 5000 configurations for the BiFeO$_3$ system. The fitted model accurately estimated the Neel temperature at 650 K, in good agreement with experimental data. Additionally, the authors investigated different types of domain walls (DWs) with characteristic angles of 109$^{\circ}$, 180$^{\circ}$, and 71$^{\circ}$. The calculated DW energies were found to be 65 J/m$^2$, 170 J/m$^2$, and 185 J/m$^2$, respectively, consistent with previous studies. Later, SpinGNN++ was developed \cite{yu2022_spinGNN++}. In addition to non-collinear magnetism, SpinGNN++ method incorporates spin-orbit coupling in explicit Heisenberg, Dzyaloshinskii-Moriya, Kitaev, biquadratic, and implicit high-order spin-lattice interactions. Monolayers of the CrTe$_2$ and CrI$_3$ systems were investigated with SpinGNN++. The training sets included 5661 and 4900 configurations, respectively. The fitted model estimated the Curie temperature for CrTe$_2$ and CrI$_3$ at 60 K and 70 K, respectively. During the investigation of the strain-induced magnetic phase diagram of the CrTe$_2$ monolayer, the authors demonstrated that when considering the spin-orbit coupling (SOC) effect, the energy of the freestanding Ferri state is 3.82 meV/atom lower than the Zigzag (ZZ) state, 6.63 meV/atom lower than the AABB state, and 20.11 meV/atom lower than the ferromagnetic (FM) state. This finding indicates that the Ferri state is the global magnetic ground state. The identification of the ferrimagnetic state in the monolayer CrTe$_2$ through the SpinGNN++ model has enhanced the understanding of the material's magnetic phase diagram and sheds light on the conflicting experimental observations of antiferromagnetic and ferromagnetic signals.

In \cite{yuan2024_magENN}, equivariant neural networks (ENNs) for magnetic materials were developed. A noteworthy aspect of this work was the inclusion of magnetic forces (i.e., negative energy gradients with respect to magnetic moments) in the training of magnetic ENNs, along with energies and forces. The authors employed the constrained DFT method described in \cite{cai2023_cDFT} to create a training set incorporating excited (non-equilibrium) magnetic states for the CrI$_3$ system. By incorporating magnetic forces into the loss function during the fitting, the authors successfully trained the magnetic ENN using fewer than 1000 configurations. The model accurately predicted various properties of CrI$_3$, including the magnon dispersion of the CrI$_3$ monolayer, energy differences between CrI$_3$ nanotubes with non-collinear magnetic moments aligned along the radial direction, and those with collinear magnetic moments. Furthermore, the authors conducted spin dynamics simulations for the twisted CrI$_3$ bilayer using the magnetic ENN. They obtained relaxed magnetic configurations and subsequently calculated the band structure of the relaxed configuration using the extended deep-learning DFT Hamiltonian method, known as xDeepH \cite{li2023_xDeepH}.

The development of magnetic MLIPs has been a significant achievement in recent years, enabling the simulation of systems with thousands of atoms, which was previously infeasible using spin-polarized DFT calculations. While many of the studies mentioned above tested the potentials on prototypical magnetic systems such as Fe, some were successfully applied to more complex multi-component systems such as BiFeO$_3$, CrTe$_2$, and CrI$_3$. Many of the developed magnetic MLIPs incorporate non-collinear magnetism. One more crucial aspect of magnetic MLIPs is the consideration of spin-orbit coupling \cite{xu2024first}, which has been accounted for in SpinGNN++ \cite{yu2022_spinGNN++}. However, a notable drawback of magnetic MLIPs is the substantial amount of DFT data required --- sometimes exceeding 10000 configurations --- to accurately fit the potentials and predict the desired properties. Efficient construction of training sets can be achieved by employing active learning (AL) strategies, similar to those previously developed for non-magnetic MLIPs, as seen in works such as \cite{frederiksen2004_bayesian_al, behler2014_al, podryabinkin2017_al, jinnouchi2019_bayesian_al, lysogorskiy2023_ace_al}.

\begin{figure}[h!]
    \centering
    \includegraphics[width=\textwidth]{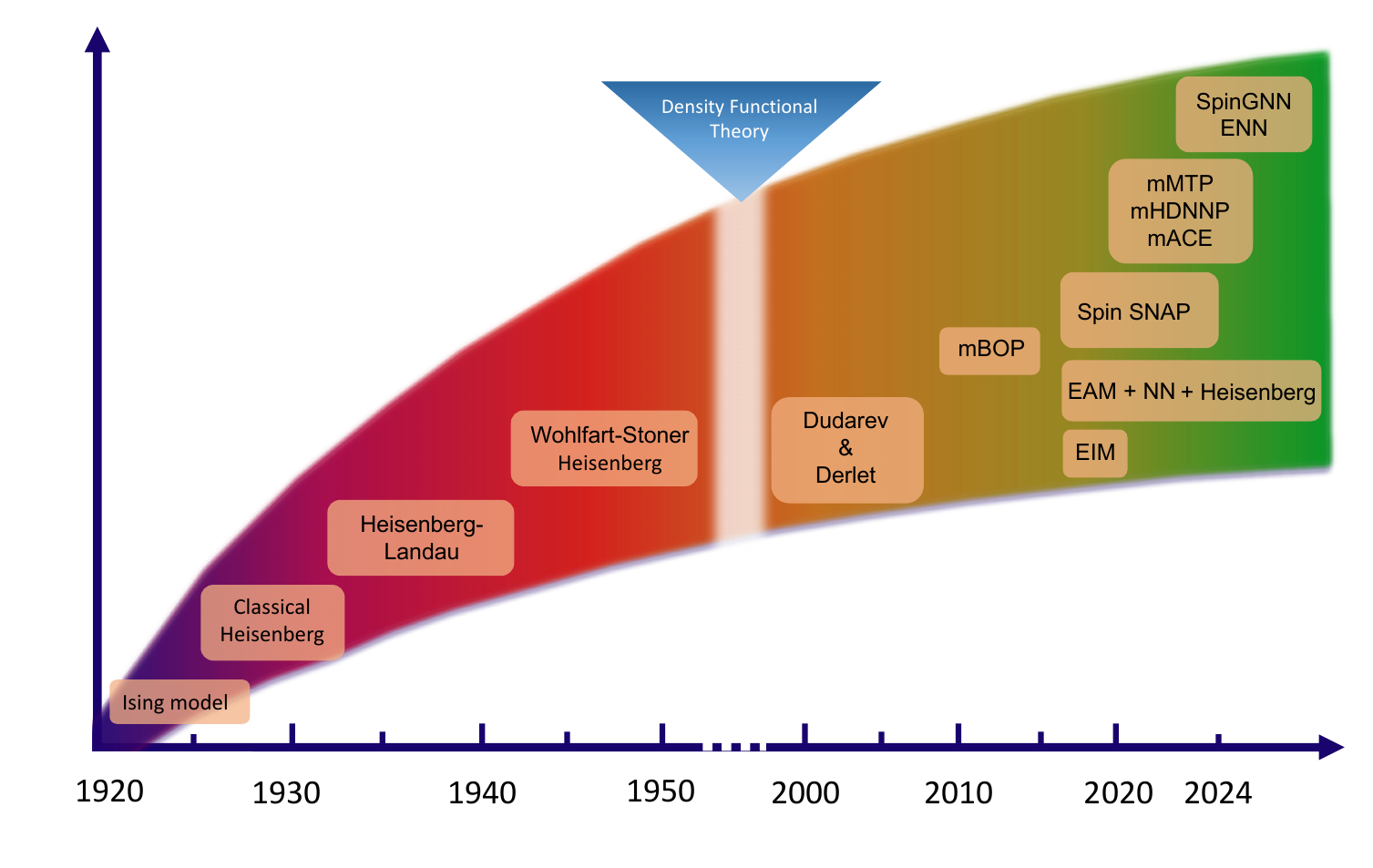}
    \caption{Timeline depicting the evolution of interatomic interaction models, ranging from the Ising model to those founded on machine learning and neural networks. This progress involves the continuous enhancement of interaction models, encompassing classical pair potentials applicable to magnetic systems, as well as Density Functional Theory. Over time, as fundamental models have been developed and machine learning methodologies have emerged, more sophisticated and precise models have been created. These encompass empirical potentials, density functionals, and machine learning techniques, enabling a more accurate representation of atomic interaction behavior in magnetic materials.
    }
    \label{fig:timeline}
\end{figure}

From the above description we conclude that significant progress has been made in recent years in the development of magnetic MLIPs. This progress builds upon previous advancements in empirical potentials, such as the Heisenberg model, which serves as the foundation for some magnetic MLIPs. Additionally, DFT calculations play a crucial role in constructing training sets for the parameterization and fitting of MLIPs. The comprehensive interatomic interaction models mentioned here are presented in Table \ref{tab:all_models}, and their timeline of development is depicted in Fig. \ref{fig:timeline}.

{\it 5. Conclusion and Future Outlook.} In this review, we have discussed various interatomic interaction models employed for the investigation of magnetic materials, including empirical interatomic potentials based on the Heisenberg model, density functional theory (DFT), and magnetic machine-learning interatomic potentials (MLIPs). These models have proven to be reliable tools for predicting properties and quantities of magnetic materials, such as the Curie temperature, Neel temperature, and magnetization. 

Recent advancements have demonstrated that magnetic MLIPs can be automatically parameterized and applied to the study of multi-component magnetic materials. These potentials offer a combination of the computational efficiency seen in empirical potentials and the accuracy of DFT calculations. As a result, magnetic MLIPs hold great promise for investigating complex multi-component materials beyond single-component or binary systems, including magnetic random alloys, perovskites, and spintronics materials.

The future development of fitting algorithms for magnetic MLIPs includes active learning techniques, which aim to select training configurations automatically and optimally, thereby reducing the need for costly DFT calculations. Moreover, an exciting direction for further progress involves the explicit incorporation of external magnetic fields within the framework of magnetic MLIPs.

In conclusion, magnetic MLIPs provide a valuable avenue for exploring the properties and behavior of magnetic materials, offering a powerful combination of computational efficiency and accuracy. The continued advancement in training algorithms and the explicit inclusion of external magnetic fields will further enhance their applicability in studying complex multi-component magnetic materials.

{\it 6. Acknowledgements.} This work was supported by Russian Science Foundation (grant number 22-73-10206, https://rscf.ru/project/22-73-10206/). We want to thank Alexey Kotyhov for fruitful discussions and Alexander Grishchenko for artistic illustrations.

\bibliographystyle{unsrt}

\begin{table}[h!]
    \centering
    \caption{Interatomic interaction models and their descriptive capabilities in atomistic simulations. A comprehensive compilation of interatomic interaction models, along with the specific properties they can accurately describe when utilized in atomistic simulations. The following abbreviations for the methods are used: Monte-Carlo (MC), mean-field (MF), spin dynamics (SD), quantum mechanics (QM), molecular dynamics (MD), spin-lattice dynamics (SLD).}
    \label{tab:all_models}
\footnotesize
 \begin{tabular}{cccccc}
        \toprule
        Year & Models & Features & Methods & \multicolumn{2}{c}{Examples of investigated} \\
        & & & & Systems & Properties \\
        \midrule
        1925 & Ising {\cite{ising1925contribution,peierls1936ising}} & collinear & MC & single-walled  & phase  \\
        {     }  & {} & & nanotube \cite{elden2022monte} & transitions (PT) \\
        & & & & ${\mathrm{Fe_7S_8}}$ \cite{Benyoussef_2020} & ${T_{\rm C}}$\\
        \midrule
        1928 & Classical Heisenberg \cite{heisenberg1985theorie} & non-collinear & MF theory & FeCo \cite{maclaren1999electronic} & order/disorder PT \\
        \midrule
        1937 & Heisenberg-Landau \cite{landau2013course} & non-collinear & MC, SD & FeRh\cite{derlet2012landau} & ${T_{\rm C}}$ \\
        {     }  & {     } & & & bcc Fe \cite{ma2012longitudinal} & PT \\
        \midrule
        1948 & Wohlfart-Stoner \cite{stoner1948mechanism} & non-collinear &  MC, MF & Fe,Ni\cite{stoner1948mechanism}, Co, Ni\cite{rosengaard1997finite} & ${T_{\rm C}}$ \\
        & Heisenberg & & & \\
        \midrule
        1964 & Density functional \cite{hohenberg_kohn_1964} & non-collinear & QM &  high-entropy & magnetic properties, ${T_{\rm C}}$ \\
        & theory (DFT)  & & & alloys \cite{ikeda2019_DFT_HEA} & phase stability, phonon spectra\\
        & & & & & elastic properties \\
        & & & & & lattice distortion \\
        \midrule
        2005 & Dudarev and Derlet \cite{dudarev2005magnetic} & non-collinear & MD, MC, SD & {bcc Fe \cite{chiesa2011optimization,eisenbach2015magnetic}} & {bulk properties} \\
         {     } & {     } & & & & {point defects} \\
         {     }  & {     } & & & {     } & {specific heat} \\
         \midrule
         2011 & magnetic Bond-order \cite{PhysRevLett.106.246402} & & {     } & bcc, fcc, hcp Fe & planar defects, ${T_{\rm C}}$\\
         {    } & potential (mBOP) & collinear & & \cite{PhysRevLett.106.246402,friak2001ab,friak2008ab} & point defects (PD) \\
         {    } &{    } &{    }& &FeCo \cite{egorov2023magnetic}& phonon spectrum \\
        {    } &{    } &{    } & &{    } & elastic properties\\
        \midrule
        2021 & Efficient interaction & non-collinear & MC & bcc, fcc Fe, Ni,  & phase diagram \\
        {    } &model (EIM)+Heisenberg \cite{li2021_eim}  &{    } &Fe-Ni \cite{li2022_eim} & PD \\
        \midrule
        2021 & magnetic High-dimensional \cite{eckhoff2021_mhdnnp} & collinear & MDMC & bcc, fcc Fe \cite{chapman2022_nn_defect_Fe} & PD, ${T_{\rm C}}$, ${T_{\rm Neel}}$ \\
         & NN potential (mHDNNP)  & & & Mn-O\cite{eckhoff2021_mhdnnp} & magnetization, ${T_{\rm Neel}}$\\
         & & & & & PD, thermal expansion\\
         \midrule
        2021 & SNAP+Heisenberg \cite{nikolov2021_snap_heisenberg} & non-collinear & MD, SD & bcc, fcc, hcp Fe \cite{nikolov2021_snap_heisenberg} & elastic constants\\
        \midrule
        2022 & spin SNAP \cite{domina2022_spinSNAP} & non-collinear & & bcc Fe \cite{domina2022_spinSNAP} & cohesive energy\\ 
        & & & & &bulk modulus \\
        \midrule
        2022 & EAM+Heisenberg-Landau+ \cite{chapman2022_nn_defect_Fe} & non-collinear & SLD & bcc and fcc Fe \cite{chapman2022_nn_defect_Fe} & cohesive energy\\
        &neural network (NN) & & & & \\
        \midrule
         2022 & magnetic Moment \cite{Novikov2022-mMTP} & collinear & MD & bcc, fcc, hcp Fe \cite{Novikov2022-mMTP} & phonon spectra\\
         & Tensor Potential (mMTP) & & & & energy/volume curve\\
         & & & & & magnetic properties\\
         \midrule
         2022 & Spin-dependent graph \cite{yu2022_spinGNN} & non-collinear & MD &$\mathrm{BiFeO_3}$ \cite{yu2022_spinGNN}&  ${T_{\rm Neel}}$ \\
          & NN (SpinGNN) & & & &domain wall (DW) \\
         \midrule
         2022 & SpinGNN++ \cite{yu2022_spinGNN++} & non-collinear & SLD & $\mathrm{CrI_3}$,  $\mathrm{CrTe_2}$\cite{yu2022_spinGNN++}&  ${T_{\rm C}}$ \\
          & & spin-orbit coupling & & &magnetic phase diagram \\
         \midrule
         2023 & multi-component mMTP \cite{Kotykhov2023-cDFT-mMTP} & collinear & relaxation & Fe-Al \cite{Kotykhov2023-cDFT-mMTP} & lattice parameter
         \\
         & & & & & magnetization\\
         \midrule
         2024 & magnetic Atomic Cluster \cite{drautz2024_noncolACE} & non-collinear & MDMC & bcc, fcc Fe\cite{drautz2024_noncolACE}& elastic constants\\
         & Expansion (mACE) & & & &  energy/volume curve\\
         & & & & & PD, phonon spectra, ${T_{\rm C}}$\\
         \midrule
        2024 & Equivariant NN (ENN) \cite{yuan2024_magENN} & non-collinear & SD  & $\mathrm{CrI_3}$ \cite{yuan2024_magENN}& magnon dispersion\\
        & & & & & band structure\\
        \bottomrule
    \end{tabular}
\end{table}

\end{document}